\documentclass[a4paper,11pt]{article}
\usepackage[T1]{fontenc} 

\makeatletter
\gdef\@fpheader{ }
\gdef\@journal{ }
\makeatother
\RequirePackage{amsmath}
\RequirePackage{amssymb}
\RequirePackage{epsfig}
\RequirePackage{graphicx}
\RequirePackage[numbers,sort&compress]{natbib}
\RequirePackage{color}
\RequirePackage[colorlinks=true
,urlcolor=blue
,anchorcolor=blue
,citecolor=blue
,filecolor=blue
,linkcolor=blue
,menucolor=blue
,pagecolor=blue
,linktocpage=true
,pdfproducer=medialab
,pdfa=true
]{hyperref}

\newif\ifnotoc\notocfalse
\newif\ifemailadd\emailaddfalse
\newif\iftoccontinuous\toccontinuousfalse
\makeatletter
\def\@subheader{\@empty}
\def\@keywords{\@empty}
\def\@abstract{\@empty}
\def\@xtum{\@empty}
\def\@dedicated{\@empty}
\def\@arxivnumber{\@empty}
\def\@collaboration{\@empty}
\def\@collaborationImg{\@empty}
\def\@proceeding{\@empty}
\def\@preprint{\@empty}

\newcommand{\subheader}[1]{\gdef\@subheader{#1}}
\newcommand{\keywords}[1]{\if!\@keywords!\gdef\@keywords{#1}\else%
\PackageWarningNoLine{\jname}{Keywords already defined.\MessageBreak Ignoring last definition.}\fi}
\renewcommand{\abstract}[1]{\gdef\@abstract{#1}}
\newcommand{\dedicated}[1]{\gdef\@dedicated{#1}}
\newcommand{\arxivnumber}[1]{\gdef\@arxivnumber{#1}}
\newcommand{\proceeding}[1]{\gdef\@proceeding{#1}}
\newcommand{\xtumfont}[1]{\textsc{#1}}
\newcommand{\correctionref}[3]{\gdef\@xtum{\xtumfont{#1} \href{#2}{#3}}}
\newcommand\jname{JHEP}
\newcommand\acknowledgments{\section*{Acknowledgments}}

\newcommand\preprint[1]{\gdef\@preprint{\hfill #1}}

\makeatother

\newcommand\note[2][]{%
\if!#1!%
\stepcounter{footnote}\footnotetext{#2}%
\else%
{\renewcommand\thefootnote{#1}%
\footnotetext{#2}}%
\fi}

\makeatletter
\newtoks\auth@toks
\renewcommand{\author}[2][]{%
  \if!#1!%
    \auth@toks=\expandafter{\the\auth@toks#2\ }%
  \else
    \auth@toks=\expandafter{\the\auth@toks#2$^{#1}$\ }%
  \fi
}
\makeatother
\makeatletter
\newtoks\affil@toks\newif\ifaffil\affilfalse
\newcommand{\affiliation}[2][]{%
\affiltrue
  \if!#1!%
    \affil@toks=\expandafter{\the\affil@toks{\item[]#2}}%
  \else
    \affil@toks=\expandafter{\the\affil@toks{\item[$^{#1}$]#2}}%
  \fi
}
\makeatother
\makeatletter
\newtoks\email@toks\newcounter{email@counter}%
\newcommand{\emailAdd}[1]{%
\emailaddtrue%
\ifnum\theemail@counter>0\email@toks=\expandafter{\the\email@toks, \@email{#1}}%
\else\email@toks=\expandafter{\the\email@toks\@email{#1}}%
\fi\stepcounter{email@counter}}
\newcommand{\@email}[1]{\href{mailto:#1}{\tt #1}}
\makeatother

\makeatletter
\newcommand*\collaboration[1]{\gdef\@collaboration{#1}}
\newcommand*\collaborationImg[2][]{\gdef\@collaborationImg{#2}}
\makeatletter
\newcommand\afterLogoSpace{\smallskip}
\newcommand\afterSubheaderSpace{\vskip3pt plus 2pt minus 1pt}
\newcommand\afterProceedingsSpace{\vskip21pt plus0.4fil minus15pt}
\newcommand\afterTitleSpace{\vskip23pt plus0.06fil minus13pt}
\newcommand\afterRuleSpace{\vskip23pt plus0.06fil minus13pt}
\newcommand\afterCollaborationSpace{\vskip3pt plus 2pt minus 1pt}
\newcommand\afterCollaborationImgSpace{\vskip3pt plus 2pt minus 1pt}
\newcommand\afterAuthorSpace{\vskip5pt plus4pt minus4pt}
\newcommand\afterAffiliationSpace{\vskip3pt plus3pt}
\newcommand\afterEmailSpace{\vskip16pt plus9pt minus10pt\filbreak}
\newcommand\afterXtumSpace{\par\bigskip}
\newcommand\afterAbstractSpace{\vskip16pt plus9pt minus13pt}
\newcommand\afterKeywordsSpace{\vskip16pt plus9pt minus13pt}
\newcommand\afterArxivSpace{\vskip3pt plus0.01fil minus10pt}
\newcommand\afterDedicatedSpace{\vskip0pt plus0.01fil}
\newcommand\afterTocSpace{\bigskip\medskip}
\newcommand\afterTocRuleSpace{\bigskip\bigskip}
\newlength{\affiliationsSep}
\newcommand\beforetochook{\pagestyle{myplain}\pagenumbering{roman}}

\DeclareFixedFont\trfont{OT1}{phv}{b}{sc}{11}

\renewcommand\maketitle{
\pagestyle{empty}
\thispagestyle{titlepage}
\setcounter{page}{0}
\noindent{\small\scshape\@fpheader}\@preprint\par

\afterLogoSpace
\if!\@subheader!\else\noindent{\trfont{\@subheader}}\fi
\afterSubheaderSpace
\if!\@proceeding!\else\noindent{\sc\@proceeding}\fi
\afterProceedingsSpace
{\LARGE\flushleft\sffamily\bfseries\@title\par}
\afterTitleSpace
\hrule height 1.5\p@%
\afterRuleSpace
\if!\@collaboration!\else
{\Large\bfseries\sffamily\raggedright\@collaboration}\par
\afterCollaborationSpace
\fi
\if!\@collaborationImg!\else
{\normalsize\bfseries\sffamily\raggedright\@collaborationImg}\par
\afterCollaborationImgSpace
\fi
{\bfseries\raggedright\sffamily\the\auth@toks\par}
\afterAuthorSpace
\ifaffil\begin{list}{}{%
\setlength{\leftmargin}{0.28cm}%
\setlength{\labelsep}{0pt}%
\setlength{\itemsep}{\affiliationsSep}%
\setlength{\topsep}{-\parskip}}
\itshape\small%
\the\affil@toks
\end{list}\fi
\afterAffiliationSpace
\ifemailadd 
\noindent\hspace{0.28cm}\begin{minipage}[l]{.9\textwidth}
\begin{flushleft}
\textit{E-mail:} \the\email@toks
\end{flushleft}
\end{minipage}
\else 
\PackageWarningNoLine{\jname}{E-mails are missing.\MessageBreak Plese use \protect\emailAdd\space macro to provide e-mails.}
\fi
\afterEmailSpace
\if!\@xtum!\else\noindent{\@xtum}\afterXtumSpace\fi
\if!\@abstract!\else\noindent{\renewcommand\baselinestretch{.9}\textsc{Abstract:}}\ \@abstract\afterAbstractSpace\fi
\if!\@keywords!\else\noindent{\textsc{Keywords:}} \@keywords\afterKeywordsSpace\fi
\if!\@arxivnumber!\else\noindent{\textsc{ArXiv ePrint:}} \href{http://arxiv.org/abs/\@arxivnumber}{\@arxivnumber}\afterArxivSpace\fi
\if!\@dedicated!\else\vbox{\small\it\raggedleft\@dedicated}\afterDedicatedSpace\fi
\ifnotoc\else
\iftoccontinuous\else\newpage\fi
\beforetochook\hrule
\tableofcontents
\afterTocSpace
\hrule
\afterTocRuleSpace
\fi
\setcounter{footnote}{0}
\pagestyle{myplain}\pagenumbering{arabic}
} 

\renewcommand{\baselinestretch}{1.1}\normalsize
\divide\@tempdima\baselineskip
\@tempcnta=\@tempdima
\skip\@mpfootins = \skip\footins
\renewcommand{\@dotsep}{10000}

\newcommand\ps@myplain{
\pagenumbering{arabic}
\renewcommand\@oddfoot{\hfill-- \thepage\ --\hfill}
\renewcommand\@oddhead{}}
\let\ps@plain=\ps@myplain

\newcommand\ps@titlepage{\renewcommand\@oddfoot{}\renewcommand\@oddhead{}}

\numberwithin{equation}{section}

\renewcommand\section{\@startsection{section}{1}{\z@}%
                                   {-3.5ex \@plus -1.3ex \@minus -.7ex}%
                                   {2.3ex \@plus.4ex \@minus .4ex}%
                                   {\normalfont\large\bfseries}}
\renewcommand\subsection{\@startsection{subsection}{2}{\z@}%
                                   {-2.3ex\@plus -1ex \@minus -.5ex}%
                                   {1.2ex \@plus .3ex \@minus .3ex}%
                                   {\normalfont\normalsize\bfseries}}
\renewcommand\subsubsection{\@startsection{subsubsection}{3}{\z@}%
                                   {-2.3ex\@plus -1ex \@minus -.5ex}%
                                   {1ex \@plus .2ex \@minus .2ex}%
                                   {\normalfont\normalsize\bfseries}}
\renewcommand\paragraph{\@startsection{paragraph}{4}{\z@}%
                                   {1.75ex \@plus1ex \@minus.2ex}%
                                   {-1em}%
                                   {\normalfont\normalsize\bfseries}}
\renewcommand\subparagraph{\@startsection{subparagraph}{5}{\parindent}%
                                   {1.75ex \@plus1ex \@minus .2ex}%
                                   {-1em}%
                                   {\normalfont\normalsize\bfseries}}

\def\fnum@figure{\textbf{\figurename\nobreakspace\thefigure}}
\def\fnum@table{\textbf{\tablename\nobreakspace\thetable}}

\long\def\@makecaption#1#2{%
  \vskip\abovecaptionskip
  \sbox\@tempboxa{\small #1. #2}%
  \ifdim \wd\@tempboxa >\hsize
    \small #1. #2\par
  \else
    \global \@minipagefalse
    \hb@xt@\hsize{\hfil\box\@tempboxa\hfil}%
  \fi
  \vskip\belowcaptionskip}

\renewenvironment{thebibliography}[1]{%
\begin{oldthebibliography}{#1}%
\small%
\raggedright%
\setlength{\itemsep}{5pt plus 0.2ex minus 0.05ex}%
}%
{%
\end{oldthebibliography}%
}

\begin{document}



\renewcommand{\thefootnote}{\fnsymbol{footnote}}

\title{\boldmath Gravitational wave scattering theory without large-distance asymptotics}%

\author[a,b,c]{Wen-Du Li,}\note{liwendu@tjnu.edu.cn}
\author[b]{Shi-Lin Li,}
\author[b]{Yu-Jie Chen,}
\author[c,b]{Yu-Zhu Chen,}\note{chenyuzhu@nankai.edu.cn}
\author[b*]{and Wu-Sheng Dai}\note{daiwusheng@tju.edu.cn.}


\affiliation[a]{College of Physics and Materials Science, Tianjin Normal University, Tianjin 300387, PR China}
\affiliation[b]{Department of Physics, Tianjin University, Tianjin 300072, P.R. China}
\affiliation[c]{Theoretical Physics Division, Chern Institute of Mathematics, Nankai University, Tianjin, 300071, P. R. China}










\abstract{
In conventional gravitational wave scattering theory, a large-distance
asymptotic approximation is employed. In this approximation, the gravitational
wave is approximated by its large-distance asymptotics. In this paper, we
establish a gravitational wave scattering theory without the large-distance
asymptotic approximation.
}

\maketitle
\flushbottom


\section{Introduction}

The gravitational wave now has been detected, the first three observations
GW150914 \cite{abbott2016observation,abbott2016astrophysical,abbott2016binary}%
, GW151226 \cite{abbott2016gw151226}, GW170104 \cite{scientific2017gw170104}
by LIGO, the observation by the advanced Virgo detector and the two advanced
LIGO detectors \cite{abbott2017gw170814}, and the gravitational wave produced
by a binary neutron star merging
\cite{abbott2017gw170817,abbott2017gravitational}. There are resent
theoretical researches on scattering of gravitational waves, such as
long-wavelength gravitational wave scattering \cite{dolan2008scattering},
rainbow scattering of gravitational plane waves \cite{stratton2019rainbow},
the Regge pole of gravitational wave scattering \cite{folacci2019regge}, and
scattering of plane-fronted gravitational waves \cite{pretorius2018black}. The
gravitational wave and the gravitational wave scattering have been studied for
many years, e.g., historically, gravitational radiations at infinity in a
asymptotically flat space \cite{torrence1967approach}, classical cross
sections \cite{westervelt1971scattering}, inelastic cross sections of
nonrotating and rotating black holes \cite{matzner1978scattering}, scattering
off a Kerr black hole \cite{handler1980gravitational}, the weak-field
gravitational scattering \cite{de1977gravitational}, gravitational wave
scattering on a Schwarzschild black hole in the low-frequency limit
\cite{matzner1977low}, scattering of small wave amplitudes and weak
gravitational fields \cite{peters1974index}, and differential cross sections
of plane gravitational waves scattering from the gravitational field of
sources in the weak-field approximation \cite{peters1976differential}. In
observation, there is also an indirect evidence for the existence of
gravitational wave \cite{hulse1975discovery}. Recently, there are many
theoretical studies on the gravitational wave scattering: the scattering and
absorption of planar gravitational waves by a Kerr black hole
\cite{dolan2008scattering}, the scattering of a low-frequency gravitational
wave by a massive compact body \cite{dolan2008scattering}, the scattering of
weak gravitational waves from a slowly rotating gravitational source
\cite{sorge2015gravitational}, the interaction of a weak gravitational wave
with matter \cite{cetoli2012interaction}, the low-energy scattering of
gravitons\ \cite{guadagnini2008graviton}, and the scattering of gravitational
waves by the weak gravitational fields of lens objects
\cite{takahashi2005scattering}.

One important approach to study gravitational waves is to study the weak-field
radiative solution of the Einstein equation. Under the weak-field
approximation, the metric $g_{\mu\nu}$ is close to the Minkowski metric
$\eta_{\mu\nu}$:%
\begin{equation}
g_{\mu\nu}=\eta_{\mu\nu}+h_{\mu\nu}%
\end{equation}
with the determinant $\left\vert h_{\mu\nu}\right\vert \ll1$
\cite{weinberg2004gravitation}. Then the Einstein equation becomes a linear
equation $\square^{2}h_{\mu\nu}=-16\pi GS_{\mu\nu}$ with a source $S_{\mu\nu}%
$. When the gravitational radiation comes in from infinity, the gravitational
wave approaches a plane wave satisfying%
\begin{equation}
\square^{2}h_{\mu\nu}=0. \label{hequation}%
\end{equation}
A plane gravitational wave impinging on a target consists of an
incident\textit{ }plane wave and a scattered wave,
\begin{equation}
h_{\mu\nu}=h_{\mu\nu}^{\text{inc}}+h_{\mu\nu}^{\text{sc}}, \label{hmunu}%
\end{equation}
where $h_{\mu\nu}^{\text{inc}}$ $=e_{\mu\nu}e^{i\mathbf{k}\cdot\mathbf{x}%
}e^{-ikt}$ is the incident wave with $e_{\mu\nu}$ the polarization tensor and
$k^{\mu}$ the wave vector and $h_{\mu\nu}^{\text{sc}}$ is the scattered wave.

In conventional gravitational wave scattering theory, a large-distance
asymptotic approximation is employed. In this approximation, the incident
plane wave $e^{i\mathbf{k\cdot x}}$ is replaced by its large-distance
asymptotics, $e^{ikr\cos\theta}\overset{r\rightarrow\infty}{\sim}\sum
_{l=0}^{\infty}\left(  2l+1\right)  P_{l}\left(  \cos\theta\right)  \frac
{1}{2ikr}\left[  e^{ikr}-\left(  -1\right)  ^{l}e^{-ikr}\right]  $, and the
scattered wave is replaced by a spherical wave, $h_{\mu\nu}^{\text{sc}%
}\overset{r\rightarrow\infty}{\sim}f_{\mu\nu}\left(  \mathbf{\hat{x}}\right)
\frac{e^{ikr}}{r}e^{-ikt}$, where $f_{\mu\nu}\left(  \mathbf{\hat{x}}\right)
$ is the scattering amplitude \cite{weinberg2004gravitation}. Consequently, in
conventional scattering theory, the gravitational wave $h_{\mu\nu}$ is
asymptotically written as
\begin{equation}
h_{\mu\nu}^{\infty}\left(  \mathbf{x},t\right)  \overset{r\rightarrow
\infty}{\sim}\left(  e_{\mu\nu}^{\infty\text{in}}e^{-ikr}+e_{\mu\nu}%
^{\infty\text{out}}e^{ikr}\right)  e^{-ikt}, \label{yinlibosbcyuan}%
\end{equation}
with%
\begin{align}
e_{\mu\nu}^{\infty\text{in}}  &  =-\frac{1}{2ikr}\sum_{l=0}^{\infty}e_{\mu\nu
}\left(  2l+1\right)  \left(  -1\right)  ^{l}P_{l}\left(  \cos\theta\right)
,\label{einfin}\\
e_{\mu\nu}^{\infty\text{out}}  &  =\frac{1}{2ikr}\sum_{l=0}^{\infty}e_{\mu\nu
}\left(  2l+1\right)  P_{l}\left(  \cos\theta\right)  +f_{\mu\nu}\left(
\mathbf{\hat{x}}\right)  \frac{1}{r}, \label{einfout}%
\end{align}
where the superscript $\infty$ denotes the large-distance asymptotics. All of
the observable quantities are described by the scattering amplitude $f_{\mu
\nu}\left(  \mathbf{\hat{x}}\right)  $ which is independent of the distance
$r$.

The large-distance asymptotic approximation, however, losses the information
of the distance. In this paper, we establish a rigorous gravitational wave
scattering theory without the large-distance asymptotic approximation.

In section \ref{Scattering theory}, we establish a gravitational wave
scattering theory without large-distance asymptotics. In section
\ref{asymptotics}, we show how this scattering theory recovers conventional
scattering theory when taking the large-distance asymptotic approximation. In
section \ref{source}, we consider the gravitational wave with a source. The
conclusion is given in section \ref{Conclusion}.

\section{Scattering theory without large-distance asymptotics
\label{Scattering theory}}

In this section, we provide an expression of $h_{\mu\nu}$ without the
large-distance asymptotic approximation.

\subsection{Gravitational wave}

\textit{Incident plane wave.} The incident\textit{ }plane wave $e^{i\mathbf{k}%
\cdot\mathbf{x}}$ can be exactly expanded as $e^{i\mathbf{k}\cdot\mathbf{x}%
}=e^{ikr\cos\theta}=\sum_{l=0}^{\infty}\left(  2l+1\right)  i^{l}j_{l}\left(
kr\right)  P_{l}\left(  \cos\theta\right)  $. In the large-distance asymptotic
approximation, the Bessel function $j_{l}\left(  kr\right)  $ is approximately
replaced by its large-distance asymptotics $j_{l}\left(  z\right)
\overset{z\rightarrow\infty}{\sim}\frac{i^{-l}}{2iz}\left[  e^{iz}-\left(
-1\right)  ^{l}e^{-iz}\right]  $. In the following, we show that, by use of
the relations $j_{l}\left(  z\right)  =\frac{1}{2}\left(  h_{l}^{\left(
1\right)  }\left(  z\right)  +h_{l}^{\left(  2\right)  }\left(  z\right)
\right)  $ and $h_{l}^{\left(  1,2\right)  }\left(  z\right)  =i^{\mp l}%
\frac{e^{\pm iz}}{\pm iz}y_{l}\left(  \mp\frac{1}{iz}\right)  $, where
$h_{l}^{\left(  1\right)  }\left(  z\right)  $ and $h_{l}^{\left(  2\right)
}\left(  z\right)  $ are the spherical Hankel functions of first and second
kinds and $y_{l}\left(  z\right)  =\sum_{k=0}^{l}\frac{\left(  l+k\right)
!}{k!\left(  l-k\right)  !}\left(  \frac{z}{2}\right)  ^{k}$ is the Bessel
polynomial \cite{liu2014scattering,li2016scattering,zhou2018acoustic}, the
expansion of the plane wave can be exactly rewritten as
\begin{equation}
e^{i\mathbf{k}\cdot\mathbf{x}}=\frac{e^{-ikr}}{-2ikr}\sum_{l=0}^{\infty
}\left(  2l+1\right)  \left(  -1\right)  ^{l}y_{l}\left(  \frac{1}%
{ikr}\right)  P_{l}\left(  \cos\theta\right)  +\frac{e^{ikr}}{2ikr}\sum
_{l=0}^{\infty}\left(  2l+1\right)  y_{l}\left(  -\frac{1}{ikr}\right)
P_{l}\left(  \cos\theta\right)  . \label{pwexpd}%
\end{equation}
The first term here describes the ingoing wave propagating inward along the
radial direction and the second term describes the outgoing wave propagating
outward along the radial direction.

The information of the distance is preserved since there is no large-distance
asymptotic approximation.

\textit{Scattered wave. }In the large-distance asymptotic approximation, the
scattered wave is approximated as a spherical wave. Nevertheless, the exact
scattered wave at a finite distance is not spherical. The scattered wave
$h_{\mu\nu}^{\text{sc}}$ without large-distance asymptotics can be written as
\cite{li2016scattering}%
\begin{equation}
h_{\mu\nu}^{\text{sc}}=\sum_{l=0}^{\infty}a_{l\mu\nu}\left(  \theta\right)
h_{l}^{\left(  1\right)  }\left(  kr\right)  e^{-ikt}, \label{hsc}%
\end{equation}
where $a_{l\mu\nu}\left(  \theta\right)  $ is the partial wave scattering amplitude.

\textit{Gravitational wave. }The incident plane wave and the scattered wave
are now expressed without the large-distance asymptotic approximation. Then
the gravitational wave (\ref{hmunu}) can be expressed as
\begin{equation}
h_{\mu\nu}=\left(  e_{\mu\nu}^{\text{in}}e^{-ikr}+e_{\mu\nu}^{\text{out}%
}e^{ikr}\right)  e^{-ikt}, \label{sbcinout}%
\end{equation}
where%
\begin{align}
e_{\mu\nu}^{\text{in}}  &  =-\frac{1}{2ikr}\sum_{l=0}^{\infty}e_{\mu\nu
}\left(  2l+1\right)  \left(  -1\right)  ^{l}y_{l}\left(  \frac{1}%
{ikr}\right)  P_{l}\left(  \cos\theta\right)  ,\label{yinliboein}\\
e_{\mu\nu}^{\text{out}}  &  =\frac{1}{2ikr}\sum_{l=0}^{\infty}\left[
e_{\mu\nu}\left(  2l+1\right)  P_{l}\left(  \cos\theta\right)  +2\left(
-i\right)  ^{l}a_{l\mu\nu}\left(  \theta\right)  \right]  y_{l}\left(
-\frac{1}{ikr}\right)  . \label{yinliboeout}%
\end{align}

Comparing Eqs. (\ref{yinlibosbcyuan}) and (\ref{sbcinout}), we can see that
the influence of the distance $r$ is embodied in the Bessel polynomial in Eq.
(\ref{pwexpd}). The gravitational wave $h_{\mu\nu}$ here depends on the
distance $r$ and when $r\rightarrow\infty$ it recovers the result given by
conventional scattering theory.

\subsection{Power of gravitational wave}

The power of the gravitational wave emitting out of a sphere of radius $r$ is
$P_{\text{out}}=\int\left\langle t_{0i}^{\text{out}}\right\rangle r^{2}\hat
{x}_{i}d\Omega$, where $\left\langle t_{\mu\nu}\right\rangle =\frac{k_{\mu
}k_{\nu}}{16\pi G}\left(  e^{\lambda\rho\ast}e_{\lambda\rho}-\frac{1}%
{2}\left\vert e_{\text{ }\lambda}^{\lambda}\right\vert ^{2}\right)  $ is the
average energy-momentum \cite{weinberg2004gravitation}. Then the power%
\begin{equation}
P_{\text{out}}=\frac{k^{2}r^{2}}{16\pi G}\int d\Omega\left[  \left(
e^{\text{out}}\right)  ^{\ast\mu\nu}\left(  e^{\text{out}}\right)  _{\mu\nu
}-\frac{1}{2}\left\vert \left(  e^{\text{out}}\right)  _{\text{ }\mu}^{\mu
}\right\vert ^{2}\right]  . \label{yinlibotout}%
\end{equation}

By Eq. (\ref{yinliboeout}) we can calculate the terms in Eq.
(\ref{yinlibotout}), respectively:%
\begin{align}
\left(  e^{\text{out}}\right)  ^{\ast\mu\nu}\left(  e^{\text{out}}\right)
_{\mu\nu}  &  =e^{\mu\nu\ast}e_{\mu\nu}\left\vert h_{l}^{\left(  1\right)
}\left(  kr\right)  \right\vert ^{2}\nonumber\\
&  +\frac{1}{k^{2}r^{2}}\sum_{l=0}^{\infty}\sum_{l^{\prime}=0}^{\infty}\left(
2l+1\right)  P_{l}\left(  \cos\theta\right)  \operatorname{Im}\left[  \left(
-i\right)  ^{l^{\prime}-1}e^{\mu\nu\ast}a_{l^{\prime}\mu\nu}\left(
\theta\right)  y_{l}\left(  \frac{1}{ikr}\right)  y_{l^{\prime}}\left(
-\frac{1}{ikr}\right)  \right] \nonumber\\
&  +\frac{1}{k^{2}r^{2}}\sum_{l=0}^{\infty}\sum_{l^{\prime}=0}^{\infty
}i^{l-l^{\prime}}a_{l}^{\mu\nu\ast}\left(  \theta\right)  a_{l^{\prime}\mu\nu
}\left(  \theta\right)  y_{l}\left(  \frac{1}{ikr}\right)  y_{l^{\prime}%
}\left(  -\frac{1}{ikr}\right)  , \label{ese}%
\end{align}%
\begin{align}
\left\vert \left(  e^{\text{out}}\right)  _{\text{ }\mu}^{\mu}\right\vert
^{2}  &  =\left\vert e_{\text{ }\mu}^{\mu}h_{l}^{\left(  1\right)  }\left(
kr\right)  \right\vert ^{2}+\frac{1}{k^{2}r^{2}}\left\vert \sum_{l=0}^{\infty
}\left(  -i\right)  ^{l}a_{l\text{ }\mu}^{\mu}\left(  \theta\right)
y_{l}\left(  -\frac{1}{ikr}\right)  \right\vert ^{2}\nonumber\\
&  +\frac{1}{k^{2}r^{2}}\sum_{l=0}^{\infty}\sum_{l^{\prime}=0}^{\infty}\left(
2l+1\right)  P_{l}\left(  \cos\theta\right)  \operatorname{Im}\left[  \left(
-i\right)  ^{l^{\prime}-1}e_{\text{ }\mu}^{\mu\ast}a_{l^{\prime}\mu}^{\mu
}\left(  \theta\right)  y_{l}\left(  \frac{1}{ikr}\right)  y_{l^{\prime}%
}\left(  -\frac{1}{ikr}\right)  \right]  . \label{enorm}%
\end{align}

The power $P_{\text{out}}$ includes three parts:%
\begin{equation}
P_{\text{out}}=P_{\text{sc}}+P_{\text{int}}+P_{\text{inc}},
\end{equation}
where%
\begin{align}
P_{\text{sc}}  &  =\frac{1}{16\pi G}\sum_{l=0}^{\infty}\sum_{l^{\prime}%
=0}^{\infty}i^{l-l^{\prime}}\int d\Omega a_{l}^{\mu\nu\ast}\left(
\theta\right)  a_{l^{\prime}\mu\nu}\left(  \theta\right)  y_{l}\left(
\frac{1}{ikr}\right)  y_{l^{\prime}}\left(  -\frac{1}{ikr}\right) \nonumber\\
&  -\frac{1}{32\pi G}\int d\Omega\left\vert \sum_{l=0}^{\infty}\left(
-i\right)  ^{l}a_{l\text{ }\mu}^{\mu}\left(  \theta\right)  y_{l}\left(
-\frac{1}{ikr}\right)  \right\vert ^{2} \label{yinlibopsc}%
\end{align}
describes the contribution of the scattering wave,%
\begin{equation}
P_{\text{inc}}=\frac{k^{2}r^{2}}{16\pi G}\int d\Omega\left(  e^{\mu\nu\ast
}e_{\mu\nu}\left\vert h_{l}^{\left(  1\right)  }\left(  kr\right)  \right\vert
^{2}+\left\vert e_{\text{ }\mu}^{\mu}h_{l}^{\left(  1\right)  }\left(
kr\right)  \right\vert ^{2}\right)
\end{equation}
describes the contribution of the incident plane wave, and%
\begin{align}
P_{\text{int}}  &  =\frac{1}{16\pi G}\sum_{l=0}^{\infty}\sum_{l^{\prime}%
=0}^{\infty}\left(  2l+1\right) \nonumber\\
&  \times\int d\Omega P_{l}\left(  \cos\theta\right)  \operatorname{Im}\left[
\left(  -i\right)  ^{l^{\prime}-1}\left(  e^{\mu\nu\ast}a_{l^{\prime}\mu\nu
}\left(  \theta\right)  -\frac{1}{2}e_{\text{ }\mu}^{\mu\ast}a_{l^{\prime}\mu
}^{\mu}\left(  \theta\right)  \right)  y_{l}\left(  \frac{1}{ikr}\right)
y_{l^{\prime}}\left(  -\frac{1}{ikr}\right)  \right]  \label{yinlibopint}%
\end{align}
describes the contribution of the interference between the incident plane wave
and the scattered wave.

Noting that a quantity containing $e_{\mu\nu}$ is the contribution from the
incident wave, a quantity containing $a_{l\mu\nu}\left(  \theta\right)  $ is
the contribution from the scattered wave, and a quantity containing the cross
term of $e_{\mu\nu}$ and $a_{l\mu\nu}\left(  \theta\right)  $ is the
contribution from the interference.

\subsection{Scattering amplitude}

In the above, the gravitational wave (\ref{hmunu}) is represented by the
partial wave scattering amplitude $a_{l\mu\nu}\left(  \theta\right)  $\ in Eq.
(\ref{sbcinout}). In order to compare the exact result, Eq. (\ref{sbcinout}),
with the asymptotic approximation given by conventional scattering theory, Eq.
(\ref{yinlibosbcyuan}), we rewrite the exact result, Eq. (\ref{hmunu}), as
\begin{equation}
h_{\mu\nu}=e_{\mu\nu}e^{i\mathbf{k}\cdot\mathbf{x}}e^{-ikt}+f_{\mu\nu}\left(
r,\theta\right)  \frac{e^{ikr}}{r}e^{-ikt},
\end{equation}
where $f_{\mu\nu}\left(  r,\theta\right)  $ plays the role of the scattering
amplitude in conventional scattering theory. By Eq. (\ref{hsc}) and the
relation $h_{l}^{\left(  1,2\right)  }\left(  kr\right)  =i^{\mp l}%
\frac{e^{\pm ikr}}{\pm ikr}y_{l}\left(  \mp\frac{1}{ikr}\right)  $, we obtain%
\begin{equation}
f_{\mu\nu}\left(  r,\theta\right)  =\frac{1}{ik}\sum_{l=0}^{\infty}\left(
-i\right)  ^{l}a_{l\mu\nu}\left(  \theta\right)  y_{l}\left(  -\frac{1}%
{ikr}\right)  . \label{frtheta}%
\end{equation}

Strictly speaking, $f_{\mu\nu}\left(  r,\theta\right)  $ here is not the
scattering amplitude like its analogue $f_{\mu\nu}\left(  \mathbf{\hat{x}%
}\right)  $\ in conventional scattering theory. In the rigorous scattering
theory, there is only the partial wave scattering amplitude $a_{l\mu\nu
}\left(  \theta\right)  $ \cite{li2016scattering}.

\section{Large-distance asymptotics \label{asymptotics}}

The rigorous result recovers the result given by conventional gravitational
wave scattering theory when $r\rightarrow\infty$. In the rigorous
gravitational wave scattering theory, the scattering is described by a series
of partial wave scattering amplitudes given by Eq. (\ref{hsc}).\ We now show
that when taking large-distance asymptotics, the scattered wave reduces to a
spherical wave and $f_{\mu\nu}\left(  r,\theta\right)  $ given by Eq.
(\ref{frtheta}) reduces to the conventional scattering amplitude $f_{\mu\nu
}\left(  \mathbf{\hat{x}}\right)  $.

\subsection{Scattering amplitude}

For $r\rightarrow\infty$, the ingoing wave (\ref{yinliboein}) asymptotically
reduces to%
\begin{align}
&  e_{\mu\nu}^{\text{in}}\overset{r\rightarrow\infty}{\sim}-\frac{e_{\mu\nu}%
}{2ikr}\sum_{l=0}^{\infty}\left(  2l+1\right)  \left(  -1\right)  ^{l}%
P_{l}\left(  \cos\theta\right) \nonumber\\
&  =-\frac{e_{\mu\nu}}{ikr}\delta\left(  1+\cos\theta\right)  \label{eininf}%
\end{align}
and the outgoing wave (\ref{yinliboeout}) asymptotically reduces to%
\begin{align}
e_{\mu\nu}^{\text{out}}  &  \overset{r\rightarrow\infty}{\sim}\frac{1}%
{2ikr}\sum_{l=0}^{\infty}\left[  \left(  2l+1\right)  P_{l}\left(  \cos
\theta\right)  e_{\mu\nu}+2\left(  -i\right)  ^{l}a_{l\mu\nu}\left(
\theta\right)  \right] \nonumber\\
&  =\frac{e_{\mu\nu}}{ikr}\delta\left(  1-\cos\theta\right)  +\frac{1}%
{ikr}\sum_{l=0}^{\infty}\left(  -i\right)  ^{l}a_{l\mu\nu}\left(
\theta\right)  ,
\end{align}
where the relation $\sum_{l=0}^{\infty}\left(  2l+1\right)  P_{l}\left(
x\right)  =2\delta\left(  1-x\right)  $ \cite{olver2010nist} is used.

Consequently, the gravitational wave (\ref{sbcinout}) when $r\rightarrow
\infty$ reduces to%
\begin{equation}
h_{\mu\nu}=\left\{  -\frac{e_{\mu\nu}}{ikr}\delta\left(  1+\cos\theta\right)
e^{-ikr}+\left[  \frac{e_{\mu\nu}}{ikr}\delta\left(  1-\cos\theta\right)
+\frac{1}{ikr}\sum_{l=0}^{\infty}\left(  -i\right)  ^{l}a_{l\mu\nu}\left(
\theta\right)  \right]  e^{ikr}\right\}  e^{-ikt}.
\end{equation}
Comparing with the asymptotic gravitational wave in conventional scattering
theory, Eq. (\ref{yinlibosbcyuan}), we obtain the scattering amplitude in
conventional scattering theory:
\begin{equation}
f_{\mu\nu}\left(  \mathbf{\hat{x}}\right)  =\frac{1}{ik}\sum_{l=0}^{\infty
}\left(  -i\right)  ^{l}a_{l\mu\nu}\left(  \theta\right)  .
\label{yinlibozhenfu}%
\end{equation}
It can be seen directly from Eq. (\ref{frtheta}) that\ $f_{\mu\nu}\left(
\mathbf{\hat{x}}\right)  $ is just the large-distance asymptotics of
$f_{\mu\nu}\left(  r,\theta\right)  $, i.e., $f_{\mu\nu}\left(  r,\theta
\right)  \overset{r\rightarrow\infty}{\sim}f_{\mu\nu}\left(  \mathbf{\hat{x}%
}\right)  $.

\subsection{Power of gravitational wave}

Similarly, the large-distance asymptotics of the powers $P_{\text{sc}}$ and
$P_{\text{int}}$ can be obtained directly.

The large-distance asymptotics of the scattering part of the power given by
Eq. (\ref{yinlibopsc}) reads
\begin{equation}
P_{\text{sc}}\overset{r\rightarrow\infty}{\sim}\frac{1}{16\pi G}\sum
_{l=0}^{\infty}\sum_{l^{\prime}=0}^{\infty}i^{l}\left(  -i\right)
^{l^{\prime}}\int d\Omega\left(  a_{l}^{\mu\nu\ast}\left(  \theta\right)
a_{l^{\prime}\mu\nu}\left(  \theta\right)  -\frac{1}{2}a_{l\text{ }\mu}%
^{\mu\ast}\left(  \theta\right)  a_{l^{\prime}\mu}^{\mu}\left(  \theta\right)
\right)  \label{ylbpscyuan}%
\end{equation}
and the large-distance asymptotics of the interference part of the power given
by Eq. (\ref{yinlibopint}) reads%
\begin{equation}
P_{\text{int}}\overset{r\rightarrow\infty}{\sim}\frac{1}{16\pi G}\sum
_{l=0}^{\infty}\sum_{l^{\prime}=0}^{\infty}\left(  2l+1\right)  \int d\Omega
P_{l}\left(  \cos\theta\right)  \operatorname{Im}\left[  \left(  -i\right)
^{l^{\prime}-1}\left(  e^{\mu\nu\ast}a_{l^{\prime}\mu\nu}\left(
\theta\right)  -\frac{1}{2}e_{\text{ }\mu}^{\mu\ast}a_{l^{\prime}\mu}^{\mu
}\left(  \theta\right)  \right)  \right]  . \label{ylbpintyuan}%
\end{equation}

Using Eq. (\ref{yinlibozhenfu}), we can rewrite Eqs. (\ref{ylbpscyuan}) and
(\ref{ylbpintyuan}) as%
\begin{align}
&  P_{\text{sc}}\overset{r\rightarrow\infty}{\sim}\frac{k^{2}}{16\pi G}\int
d\Omega\left(  \frac{1}{-ik}\sum_{l=0}^{\infty}i^{l}a_{l}^{\mu\nu\ast}\left(
\theta\right)  \frac{1}{ik}\sum_{l^{\prime}=0}^{\infty}\left(  -i\right)
^{l^{\prime}}a_{l^{\prime}\mu\nu}\left(  \theta\right)  \right.  -\left.
\frac{1}{2}\frac{1}{-ik}\sum_{l=0}^{\infty}i^{l}a_{l\text{ }\mu}^{\mu\ast
}\left(  \theta\right)  \frac{1}{ik}\sum_{l^{\prime}=0}^{\infty}\left(
-i\right)  ^{l^{\prime}}a_{l^{\prime}\mu}^{\mu}\left(  \theta\right)  \right)
\nonumber\\
&  =\frac{k^{2}}{16\pi G}\int d\Omega\left(  f^{\lambda\rho\ast}\left(
\mathbf{\hat{x}}\right)  f_{\lambda\rho}\left(  \mathbf{\hat{x}}\right)
-\frac{1}{2}\left\vert f_{\text{ }\mu}^{\mu}\left(  \mathbf{\hat{x}}\right)
\right\vert ^{2}\right)  , \label{ylbpscyuan2}%
\end{align}%
\begin{align}
&  P_{\text{int}}\overset{r\rightarrow\infty}{\sim}\frac{k^{2}}{16\pi G}\int
d\Omega\operatorname{Im}\left[  -\frac{e^{\mu\nu\ast}}{k}\sum_{l=0}^{\infty
}\left(  2l+1\right)  P_{l}\left(  \cos\theta\right)  \frac{1}{ik}%
\sum_{l^{\prime}=0}^{\infty}\left(  -i\right)  ^{l^{\prime}}a_{l^{\prime}%
\mu\nu}\left(  \theta\right)  \right. \nonumber\\
&  +\left.  \frac{1}{2}\frac{e_{\text{ }\mu}^{\mu\ast}}{k}\sum_{l=0}^{\infty
}\left(  2l+1\right)  P_{l}\left(  \cos\theta\right)  \frac{1}{ik}%
\sum_{l^{\prime}=0}^{\infty}\left(  -i\right)  ^{l^{\prime}}a_{l^{\prime}\mu
}^{\mu}\left(  \theta\right)  \right] \nonumber\\
&  =\frac{k^{2}}{8\pi G}\int d\Omega\left(  1-\mathbf{\hat{k}\cdot\hat{x}%
}\right)  \operatorname{Im}\left\{  -\frac{1}{k}\left[  e^{\mu\nu\ast}%
f_{\mu\nu}\left(  \mathbf{\hat{x}}\right)  -\frac{1}{2}e_{\text{ }\mu}%
^{\mu\ast}f_{\mu}^{\mu}\left(  \mathbf{\hat{x}}\right)  \right]  \right\}  ,
\label{ylbpintyuan2}%
\end{align}
where $\sum_{l=0}^{\infty}\left(  2l+1\right)  P_{l}\left(  x\right)
=2\delta\left(  1-x\right)  $ with $\cos\theta=\mathbf{\hat{k}\cdot\hat{x}}$
is used.

These are just the powers given by conventional gravitational wave scattering
theory \cite{weinberg2004gravitation}. The leading modification of the powers
to conventional scattering theory is%
\begin{equation}
\Delta P_{\text{sc}}=\frac{1}{64\pi Gk^{2}r^{2}}\sum_{l=0}^{\infty}%
\sum_{l^{\prime}=0}^{\infty}l\left(  l+1\right)  l^{\prime}\left(  l^{\prime
}+1\right)  i^{l-l^{\prime}}\int d\Omega\left(  a_{l}^{\mu\nu\ast}\left(
\theta\right)  a_{l^{\prime}\mu\nu}\left(  \theta\right)  -\frac{1}%
{2}a_{l\text{ }\mu}^{\mu\ast}\left(  \theta\right)  a_{l^{\prime}\mu}^{\mu
}\left(  \theta\right)  \right)  ,
\end{equation}%
\begin{align}
\Delta P_{\text{int}}  &  =\frac{1}{64\pi Gk^{2}r^{2}}\sum_{l=0}^{\infty}%
\sum_{l^{\prime}=0}^{\infty}\left(  2l+1\right)  l\left(  l+1\right)
l^{\prime}\left(  l^{\prime}+1\right) \nonumber\\
&  \times\int d\Omega P_{l}\left(  \cos\theta\right)  \operatorname{Im}\left[
\left(  -i\right)  ^{l^{\prime}-1}\left(  e^{\mu\nu\ast}a_{l^{\prime}\mu\nu
}\left(  \theta\right)  -\frac{1}{2}e_{\text{ }\mu}^{\mu\ast}a_{l^{\prime}\mu
}^{\mu}\left(  \theta\right)  \right)  \right]  .
\end{align}

\section{Gravitational wave with source \label{source}}

\subsection{Gravitational wave}

In the above, we consider the gravitational wave without sources, which is
determined by the homogeneous equation (\ref{hequation}). This result can be
directly applied to the gravitational wave with sources.

A gravitational wave with a source is determined by the inhomogeneous equation%
\begin{equation}
\square^{2}h_{\mu\nu}^{\text{source}}=-16\pi GS_{\mu\nu}, \label{sourceeq}%
\end{equation}
where the source
\begin{equation}
S_{\mu\nu}=T_{\mu\nu}-\frac{1}{2}\eta_{\mu\nu}T_{\text{ }\lambda}^{\lambda}
\label{Smunu}%
\end{equation}
with $T_{\mu\nu}$ the energy-momentum tensor.

The inhomogeneous equation (\ref{sourceeq}) has a retarded potential
solution:
\begin{equation}
h_{\mu\nu}^{\text{source}}\left(  \mathbf{x},t\right)  =4G\int d^{3}%
\mathbf{x}^{\prime}\frac{S_{\mu\nu}\left(  \mathbf{x}^{\prime},t-\left\vert
\mathbf{x}-\mathbf{x}^{\prime}\right\vert \right)  }{\left\vert \mathbf{x}%
-\mathbf{x}^{\prime}\right\vert }. \label{rp}%
\end{equation}

Furthermore, a solution of the inhomogeneous equation (\ref{sourceeq}) plus a
solution of the homogeneous equation (\ref{hequation}) (under the harmonic
coordinate condition $\frac{\partial}{\partial x^{\mu}}h_{\text{ }\nu}^{\mu
}=\frac{1}{2}\frac{\partial}{\partial x^{\nu}}h_{\text{ }\mu}^{\mu}$), which
describes the gravitational wave at infinity, is still a solution of the
inhomogeneous equation (\ref{sourceeq}).

To apply the result of gravitational waves without sources to gravitational
waves with sources, we Fourier expand the source $T_{\mu\nu}\left(
\mathbf{x},t\right)  $ as%
\begin{equation}
T_{\mu\nu}\left(  \mathbf{x},t\right)  =\sum_{k}T_{\mu\nu}\left(
\mathbf{x},k\right)  e^{-ikt}. \label{FExpand}%
\end{equation}

First consider a source with only one Fourier component, i.e.,%
\begin{equation}
T_{\mu\nu}=T_{\mu\nu}\left(  \mathbf{x},k\right)  e^{-ikt}. \label{oneterm}%
\end{equation}
Then the gravitational wave with the source (\ref{oneterm}), by Eq.
(\ref{rp}), reads%
\begin{equation}
h_{\mu\nu}^{\text{source}}\left(  \mathbf{x},t\right)  =4G\int d^{3}%
\mathbf{x}^{\prime}\frac{S_{\mu\nu}\left(  \mathbf{x}^{\prime},k\right)
}{\left\vert \mathbf{x}-\mathbf{x}^{\prime}\right\vert }e^{-ikt+ik\left\vert
\mathbf{x}-\mathbf{x}^{\prime}\right\vert }.
\end{equation}
For $r=\left\vert \mathbf{x}\right\vert \rightarrow\infty$, we have
$\left\vert \mathbf{x}-\mathbf{x}^{\prime}\right\vert \sim r-\mathbf{x}%
^{\prime}\cdot\frac{\mathbf{x}}{r}$. We then arrive at%
\begin{equation}
h_{\mu\nu}^{\text{source}}\left(  \mathbf{x},t\right)  \sim e^{-ikt+ikr}%
\left[  4G\int d^{3}\mathbf{x}^{\prime}\frac{S_{\mu\nu}\left(  \mathbf{x}%
^{\prime},k\right)  }{r}e^{-ik\mathbf{x}^{\prime}\cdot\frac{\mathbf{x}}{r}%
}\right]  .
\end{equation}
For large $kr$, this gravitational wave with the source (\ref{oneterm}) can be
regarded as a plane wave%
\begin{equation}
h_{\mu\nu}^{\text{source}}\left(  \mathbf{x},t\right)  =e_{\mu\nu}\left(
\mathbf{x,}t\right)  e^{-ikt+ikr} \label{hmunue}%
\end{equation}
with the polarization tensor%
\begin{equation}
e_{\mu\nu}\left(  \mathbf{x},t\right)  =4G\int d^{3}\mathbf{x}^{\prime}%
\frac{S_{\mu\nu}\left(  \mathbf{x}^{\prime},k\right)  }{r}e^{-ik\mathbf{x}%
^{\prime}\cdot\frac{\mathbf{x}}{r}}.
\end{equation}

Now we consider an arbitrary source.

Fourier expanding the source $T_{\mu\nu}\left(  \mathbf{x},t\right)  $ as Eq.
(\ref{FExpand}), by Eq. (\ref{Smunu}), we then have the Fourier expansion of
$S_{\mu\nu}\left(  \mathbf{x},t\right)  $:
\begin{equation}
S_{\mu\nu}\left(  \mathbf{x},t\right)  =\sum_{k}S_{\mu\nu}\left(
\mathbf{x},k\right)  e^{-ikt}.
\end{equation}
Then the gravitational wave with the source $T_{\mu\nu}\left(  \mathbf{x}%
,t\right)  $, by Eq. (\ref{rp}), reads%
\begin{equation}
h_{\mu\nu}^{\text{source}}\left(  \mathbf{x},t\right)  =4G\sum_{k}\int
d^{3}\mathbf{x}^{\prime}\frac{S_{\mu\nu}\left(  \mathbf{x}^{\prime},k\right)
}{\left\vert \mathbf{x}-\mathbf{x}^{\prime}\right\vert }e^{-ikt+ik\left\vert
\mathbf{x}-\mathbf{x}^{\prime}\right\vert }. \label{hmunuSource}%
\end{equation}
Substituting Eq. (\ref{hmunue}) into Eq. (\ref{hmunuSource}) gives%
\begin{align}
h_{\mu\nu}^{\text{source}}\left(  \mathbf{x},t\right)   &  =\sum_{k}\left[
4G\int d^{3}\mathbf{x}^{\prime}\frac{S_{\mu\nu}\left(  \mathbf{x}^{\prime
},k\right)  }{r}e^{-ik\mathbf{x}^{\prime}\cdot\frac{\mathbf{x}}{r}}\right]
e^{-ikt+ikr}\nonumber\\
&  =\sum_{k}e_{\mu\nu}\left(  \mathbf{x,t}\right)  e^{-ikt+ikr}.
\label{hmunu2}%
\end{align}
By the exact result (\ref{pwexpd}), the gravitational wave with the source
$T_{\mu\nu}\left(  \mathbf{x},t\right)  $ can be expressed as%
\begin{align}
h_{\mu\nu}^{\text{source}}\left(  \mathbf{x},t\right)   &  =\sum_{k}e_{\mu\nu
}\left(  \mathbf{x,t}\right)  e^{-ikt}\frac{e^{-ikr}}{-2ikr}\sum_{l=0}%
^{\infty}\left(  2l+1\right)  \left(  -1\right)  ^{l}y_{l}\left(  \frac
{1}{ikr}\right)  P_{l}\left(  \cos\theta\right) \nonumber\\
&  +\sum_{k}e_{\mu\nu}\left(  \mathbf{x,t}\right)  e^{-ikt}\frac{e^{ikr}%
}{2ikr}\sum_{l=0}^{\infty}\left(  2l+1\right)  y_{l}\left(  -\frac{1}%
{ikr}\right)  P_{l}\left(  \cos\theta\right)  .
\end{align}
This result recovers the large-distance asymptotically approximate result in
the conventional scattering theory when $y_{l}\left(  \frac{1}{ikr}\right)
\rightarrow1$.

\subsection{Power of gravitational wave}

In this section, we consider the power of the gravitational wave with a source.

For a source $T_{\mu\nu}\left(  \mathbf{x},t\right)  $, by the incident wave
(\ref{hmunu2}), the scattering power of the gravitational wave can be obtained
by summing over the powers of the single frequency plane gravitational waves,
Eqs. (\ref{yinlibopsc}) and (\ref{yinlibopint}):%
\begin{align}
P_{\text{sc}}^{\text{source}} &  =\sum_{k}e_{\mu\nu}\left(  \mathbf{x,t}%
\right)  \left\{  \frac{1}{16\pi G}\sum_{l=0}^{\infty}\sum_{l^{\prime}%
=0}^{\infty}i^{l-l^{\prime}}\int d\Omega a_{l}^{\mu\nu\ast}\left(
\theta\right)  a_{l^{\prime}\mu\nu}\left(  \theta\right)  y_{l}\left(
\frac{1}{ikr}\right)  y_{l^{\prime}}\left(  -\frac{1}{ikr}\right)  \right.
\nonumber\\
&  \left.  -\frac{1}{32\pi G}\int d\Omega\left\vert \sum_{l=0}^{\infty}\left(
-i\right)  ^{l}a_{l\text{ }\mu}^{\mu}\left(  \theta\right)  y_{l}\left(
-\frac{1}{ikr}\right)  \right\vert ^{2}\right\}  ,
\end{align}%
\begin{align}
P_{\text{int}}^{\text{source}} &  =\sum_{k}e_{\mu\nu}\left(  \mathbf{x,t}%
\right)  \left\{  \frac{1}{16\pi G}\sum_{l=0}^{\infty}\sum_{l^{\prime}%
=0}^{\infty}\left(  2l+1\right)  \right.  \nonumber\\
&  \left.  \times\int d\Omega P_{l}\left(  \cos\theta\right)
\operatorname{Im}\left[  \left(  -i\right)  ^{l^{\prime}-1}\left(  e^{\mu
\nu\ast}a_{l^{\prime}\mu\nu}\left(  \theta\right)  -\frac{1}{2}e_{\text{ }\mu
}^{\mu\ast}a_{l^{\prime}\mu}^{\mu}\left(  \theta\right)  \right)  y_{l}\left(
\frac{1}{ikr}\right)  y_{l^{\prime}}\left(  -\frac{1}{ikr}\right)  \right]
\right\}  .
\end{align}
The modifications of the power to the conventional scattering theory are%
\begin{align}
\Delta P_{\text{sc}} &  =\sum_{k}e_{\mu\nu}\left(  \mathbf{x,t}\right)
\left\{  \frac{1}{64\pi Gk^{2}r^{2}}\sum_{l=0}^{\infty}\sum_{l^{\prime}%
=0}^{\infty}l\left(  l+1\right)  l^{\prime}\left(  l^{\prime}+1\right)
i^{l-l^{\prime}}\right.  \\
&  \left.  \times\int d\Omega\left(  a_{l}^{\mu\nu\ast}\left(  \theta\right)
a_{l^{\prime}\mu\nu}\left(  \theta\right)  -\frac{1}{2}a_{l\text{ }\mu}%
^{\mu\ast}\left(  \theta\right)  a_{l^{\prime}\mu}^{\mu}\left(  \theta\right)
\right)  \right\}  ,
\end{align}%
\begin{align}
\Delta P_{\text{int}} &  =\sum_{k}e_{\mu\nu}\left(  \mathbf{x,t}\right)
\left\{  \frac{1}{64\pi Gk^{2}r^{2}}\sum_{l=0}^{\infty}\sum_{l^{\prime}%
=0}^{\infty}\left(  2l+1\right)  l\left(  l+1\right)  l^{\prime}\left(
l^{\prime}+1\right)  \right.  \nonumber\\
&  \left.  \times\int d\Omega P_{l}\left(  \cos\theta\right)
\operatorname{Im}\left[  \left(  -i\right)  ^{l^{\prime}-1}\left(  e^{\mu
\nu\ast}a_{l^{\prime}\mu\nu}\left(  \theta\right)  -\frac{1}{2}e_{\text{ }\mu
}^{\mu\ast}a_{l^{\prime}\mu}^{\mu}\left(  \theta\right)  \right)  \right]
\right\}  .
\end{align}
The leading modification is the $p$-wave modification:%
\begin{equation}
\Delta P_{\text{sc}}^{p\text{-wave}}=\sum_{k}\frac{1}{k^{2}r^{2}}e_{\mu\nu
}\left(  \mathbf{x,t}\right)  \frac{1}{16\pi G}\int d\Omega\left(  a_{1}%
^{\mu\nu\ast}\left(  \theta\right)  a_{1\mu\nu}\left(  \theta\right)
-\frac{1}{2}a_{1\text{ }\mu}^{\mu\ast}\left(  \theta\right)  a_{1\mu}^{\mu
}\left(  \theta\right)  \right)  ,
\end{equation}%
\begin{equation}
\Delta P_{\text{int}}^{p\text{-wave}}=\sum_{k}\frac{1}{k^{2}r^{2}}e_{\mu\nu
}\left(  \mathbf{x,t}\right)  \frac{3}{16\pi G}\int d\Omega\cos\theta
\operatorname{Im}\left[  \left(  e^{\mu\nu\ast}a_{1\mu\nu}\left(
\theta\right)  -\frac{1}{2}e_{\text{ }\mu}^{\mu\ast}a_{1\mu}^{\mu}\left(
\theta\right)  \right)  \right]  .
\end{equation}

As a comparison, the $p$-wave contributions in the conventional scattering
theory, by Eqs. (\ref{ylbpscyuan}) and (\ref{ylbpintyuan}), are%
\begin{equation}
P_{\text{sc}}^{\text{conventional}}\overset{p\text{-wave}}{=}\sum_{k}e_{\mu
\nu}\left(  \mathbf{x,t}\right)  \frac{1}{16\pi G}\int d\Omega\left(
a_{1}^{\mu\nu\ast}\left(  \theta\right)  a_{1\mu\nu}\left(  \theta\right)
-\frac{1}{2}a_{1\text{ }\mu}^{\mu\ast}\left(  \theta\right)  a_{1\mu}^{\mu
}\left(  \theta\right)  \right)
\end{equation}%
\begin{equation}
P_{\text{int}}^{\text{conventional}}\overset{p\text{-wave}}{=}\sum_{k}%
e_{\mu\nu}\left(  \mathbf{x,t}\right)  \frac{3}{16\pi G}\int d\Omega\cos
\theta\operatorname{Im}\left[  \left(  e^{\mu\nu\ast}a_{l^{\prime}\mu\nu
}\left(  \theta\right)  -\frac{1}{2}e_{\text{ }\mu}^{\mu\ast}a_{l^{\prime}\mu
}^{\mu}\left(  \theta\right)  \right)  \right]  .
\end{equation}

\section{Conclusion \label{Conclusion}}

In summary, the conventional weak-field gravitational wave scattering theory
is a large-distance asymptotically approximate theory. In this paper, we
establish a rigorous gravitational wave scattering theory without the
large-distance asymptotic approximation.\ In the rigorous scattering theory,
the information of the distance is preserved.

There is also an important issue in gravitational wave scattering: the
calculation of the scattering amplitude and the partial wave scattering
amplitude. In quantum mechanical scattering theory, the scattering amplitude
is described by the scattering phase shift. We have developed a method for the
calculation of the scattering phase shift\ based on the heat kernel theory
\cite{pang2012relation,li2015heat} and the integral equation method of scalar
scattering in curved spacetime \cite{li2018scalar}. These methods can also be
used to calculate the scattering amplitude in gravitational wave scattering.


\acknowledgments

We are very indebted to Dr G. Zeitrauman for his encouragement. This work is supported in part by Special Funds for theoretical physics Research Program of the NSFC under Grant No. 11947124, Nankai Zhide foundation, and NSFC under Grant Nos. 11575125 and 11675119.










\providecommand{\href}[2]{#2}\begingroup\raggedright\endgroup


\end{document}